\documentclass[12pt,preprint]{aastex}

\begin{document}

\title{Local Tadpole Galaxies}

\author{Debra Meloy Elmegreen\altaffilmark{1},
Bruce G. Elmegreen\altaffilmark{2},
Jorge S\'anchez Almeida\altaffilmark{3},
Casiana Mu\~noz-Tu\~n\'on\altaffilmark{3},
Joseph Putko\altaffilmark{1, 4},
Janosz Dewberry\altaffilmark{1}}
\altaffiltext{1}{Vassar College, Dept. of Physics and Astronomy, Poughkeepsie, NY 12604}
\altaffiltext{2}{IBM Research Division, T.J. Watson Research Center, Yorktown Hts., NY 10598}
\altaffiltext{3}{Instituto de Astrof'sica de Canarias, C/ via L\'actea, s/n, 38205, La Laguna, Tenerife, Spain}
\altaffiltext{4}{Department of Physics, Middlebury College, Middlebury, VT 05753}

\begin{abstract}
Tadpole galaxies have a giant star-forming region at the end of an elongated
intensity distribution.  Here we use SDSS data to determine the ages, masses,
and surface densities of the heads and tails in 14 local tadpoles selected
from the Kiso and Michigan surveys of UV-bright galaxies, and we compare them
to tadpoles previously studied in the Hubble Ultra Deep Field. The young
stellar mass in the head scales linearly with restframe galaxy luminosity,
ranging from $\sim10^{5}\;M_\odot$ at galaxy absolute magnitude $U=-13$ mag
to $10^{9}\;M_{\odot}$ at $U=-20$ mag. The corresponding head surface density
increases from several $M_{\odot}$ pc$^{-2}$ locally to $10-100\; M_{\odot}$
pc$^{-2}$ at high redshift, and the star formation rate per unit area in the
head increases from $\sim0.01\;M_\odot$ yr$^{-1}$ kpc$^{-2}$ locally to
$\sim1\;M_\odot$ yr$^{-1}$ kpc$^{-2}$ at high $z$. These local values are
normal for star-forming regions, and the increases with redshift are
consistent with other cosmological star formation rates, most likely
reflecting an increase in gas abundance. The tails in the local sample look
like bulge-free galaxy disks.  Their photometric ages decrease from several
Gyr to several hundred Myr with increasing $z$, and their surface densities
are more constant than the surface densities of the heads. The far outer
intensity profiles in the local sample are symmetric and exponential. We
suggest that most local tadpoles are bulge-free galaxy disks with lopsided
star formation, perhaps from environmental effects such as ram pressure or
disk impacts, or from a Jeans length comparable to half the disk size.
\end{abstract} 

\section{Introduction}
Deep studies of galaxies over the past decade have revealed a variety
of peculiar morphologies at high redshift \citep{abraham,elm05,consel09}. One type common in the
early universe but rare today is the tadpole, first identified in the
Hubble Deep Field by \cite{vdb}. Tadpoles consist of a massive
star-forming clump at one end (the ``head'') and a long diffuse region
to one side (the ``tail''). In the Hubble Ultra Deep Field (UDF),
tadpoles constitute 10\% (97 galaxies) of all the galaxies larger than
10 pixels in diameter \citep{elm07}. They represent 6\% of UDF galaxies
identified by \cite{straughn} and \cite{wind} with an automated search
algorithm. In the UDF tadpoles we studied, the heads have stellar
masses of $10^7$ to $10^8$ $M_{\odot}$ and ages of $\sim10^8$ yrs at
redshift $z\sim2-4$ \citep{elm10}.

There are also tadpoles in the local universe. Blue compact dwarf (BCD) galaxies with tadpole shapes, sometimes
called cometary \citep{mark,loose,cairos1,cairos2}, were designated sub-class iI,c by
\cite{noeske00}. Several examples are shown in the BCD study by \cite{paz},
and two more by \cite{kniazev}. A large fraction of extremely metal-poor BCDs
(XBCDs) studied by \cite{pap} and \cite{morales} resemble tadpoles. The Kiso
survey of UV bright galaxies \citep{kiso-all} also contains tadpoles among
those classified as irregular with blue clumps (Ic) or irregular with a
single giant clump (Ig).  Most of the present paper concerns the Kiso
tadpoles, which amount to less than 0.2\% of the Kiso sample (as discussed in Section \ref{sample}).

Tadpole structure could have a variety of origins. Some tadpoles could be
edge-on disks with a single massive star-forming region at one end.
\cite{elm10} showed lopsided ring-like galaxies that would look like tadpoles
if viewed edge-on. Tadpoles could also result from mergers
\citep{straughn,wind} like the famous case of UGC 10214, which is called the
Tadpole Galaxy. Another example is II Zw 40 \citep{baldwin,brinks}. However,
\cite{campos} suggest that blue compact dwarf galaxies tend to be relatively
isolated, so the tadpole shapes among these may not generally be mergers.
\cite{pap} also suggested that XBCD tadpoles are not mergers, because of the
lack of tidal features.

Another possibility is that the lopsided starburst results from ram
compression by motion through an intergalactic medium. There are two
possibilities. First, tadpoles could be gas-poor disks with star formation
triggered on the leading side and the rest of the disk visible as a red tail
of older stars. Then the center of mass would be near the center of the
underlying red disk, which is somewhere between the head and the tail.
Alternatively, they could be heavily stripped galaxies with star formation
and old stars at the leading edge, perhaps still centered on a dark matter
core, and a tail made mostly from star formation in the stripped gas. Then
the tail would be blue. Both cases are interesting because then tadpoles
could be used as tracers of intergalactic gas. In this interpretation,
tadpoles are more common at high redshift because the intergalactic medium is
denser there \citep{elm10}.  For example, the scenario in which dwarf
Irregular galaxies are converted into dwarf spheroidal galaxies in the
gaseous halos of larger galaxies
\citep{lin83,vdb94,grebel03,mayer06,weisz11}, or in the hot intergalactic gas
of galaxy clusters \citep{boselli08,vanzee04}, could involve a tadpole phase
for the dwarf as the gas and young stars are pulled behind. IC 3418 in the
Virgo cluster may be an example of this \citep{chung09,hester10,fumagalli11}.

A third possibility is that some tadpoles are normal galaxies with a
large turbulent Jeans length for gravitational collapse of the
interstellar medium.  The ratio of the Jeans length to the galaxy size
is approximately the square of the ratio of the gas velocity dispersion
to the rotation speed. Galaxies with relatively large Jeans lengths
have big star formation clumps, and there is only room for a few of
these clumps in the disk. Local dwarf Irregulars and BCDs are like this
because of their low rotation speeds, while young massive galaxies are
like this because of their high velocity dispersions
\citep{forster06,elm09}. If there is only one clump at any particular
time, then the resulting star formation will appear lopsided. From the
right perspective, such a galaxy may look like a tadpole.

In this paper, we examine tadpole galaxies in the local universe for
comparison with high redshift tadpoles. We determine star formation rates and
surface densities in the heads and tails as a function of galaxy mass and
luminosity to see whether the star formation is bursting now or relatively
quiescent, and to see whether the tails are redder and older in the local
galaxies than at high redshift. If tadpole structure is a phase of galaxy evolution and local tadpoles are caused by the same events or processes as high redshift tadpoles, then detailed studies of the local tadpoles may help illuminate these processes at high redshift.

We discuss our sample of local tadpole galaxies in
Section \ref{sect:reduc}. Properties of their heads and tails and a
comparison with high-redshift tadpoles are discussed in Section
\ref{results}. Our conclusions are in Section \ref{sect:conc}.

\section{The Sample and Analysis}\label{sect:reduc}
\subsection{Sample selection}
\label{sample} The Kiso Survey for Ultraviolet-Excess Galaxies
\citep{kiso-all} detected several thousand UV-bright galaxies from
ultraviolet and red images taken with the Kiso Schmidt telescope. The survey
categorized galaxy morphologies into several types using the Palomar
Observatory Sky Survey and the Sloan Digital Sky Survey (SDSS; Stoughton et
al. 2002). The Kiso survey has 176 Ic galaxies and 83 Ig galaxies, amounting
to 2.6\% of the total 9908 Kiso galaxies. Of the 259 Ic and Ig types, 158 are
in SDSS: these include 111 out of the 176 Ic types, and 47 out of the 83 Ig
types. We examined these 158 galaxies on the SDSS images.  Many of the
galaxies in these Ic and Ig categories are spiral galaxies with one or more
bright star-forming regions. Others are irregulars with bright star
formation, or likely interactions and mergers. A small number are lopsided
with a bulge and inner disk that is offset from the outer disk (e.g.,
KUG0226+251B, KUG1129+144), or with bright star formation only on one side
(KUG2306+227). We chose galaxies for the present study that are so lopsided
that their brightest clump is far to one end and the rest of the galaxy is
mostly featureless, meaning that it has no bulge or other significant star
formation elsewhere. We consider these galaxies to be local tadpoles. We are
interested in measuring their properties and comparing them to high redshift
tadpoles.

Among the 158 Ic and Ig galaxies in SDSS, 95 have SDSS-DR7 spectra
\citep{aba}, and of these, 13 are good cases of tadpoles: 4 of type Ic and 9
of type Ig.  Of the remaining 63 without spectra, we identify one more
tadpole (KUG0010+371). If we consider this fraction of 14 tadpoles out of 158
Ic and Ig galaxies in SDSS to be representative, and extrapolate to the 259
Ic and Ig galaxies in Kiso, then the percentage of tadpoles among the 9908
galaxies in Kiso is $\sim0.2$\%. This is much smaller than the 10\% tadpoles
in the UDF.

Each of the 13 tadpoles in our Kiso sample with SDSS spectra shows emission lines. The Kiso
survey divides galaxies into bins of H (high), M (medium), and L (Low) for
the emission line strengths; 12 of our 13 tadpoles with spectra have high
emission and one (Kiso 6610) has medium emission. For reference, 82 of the
211 Ic and Ig galaxies fall into the H category. We also viewed all BCD types
iI,c \citep{noeske00} listed in NED\footnote{NASA/IPAC Extragalactic
Database, http://ned.ipac.caltech.edu/} that were also in SDSS. One more
tadpole with a known distance was selected, UM 417, which is from the University of Michigan IV
survey of emission line objects \citep{macalpine}. In the rest of this paper, we measure the properties of these 14  galaxies with known distances (i.e., 13 Kiso galaxies + 1 UM galaxy).

Archival images and spectra of these local tadpole galaxies were downloaded
from the SDSS survey. The images include 5 passbands (Gunn filters
$u,g,r,i,z$) taken with the 2.5m telescope at Apache Point, NM. Photometric
measurements on sky-subtracted images were done with IRAF (Image Reduction
and Analysis Facility) to determine the masses and ages of the clumps and
tails, as described in \ref{sub-phot}. The spectra have resolution 2000 and
cover an area 3$^{\prime\prime}$ in diameter in the galaxies. Given the
distances of our targets (Table \ref{tab1}),  3$^{\prime\prime}$ represents a
linear scale between 0.1 and 2.5 kpc, with a typical value of around 0.5 kpc.
These spectra were used to estimate H$\alpha$ star formation rates, as
described in Section \ref{sub-sfr}. The SDSS spectral resolution corresponds
to a velocity dispersion on the order of 70 km s$^{-1}$. The H$\alpha$ fluxes
are independent of this dispersion, because they are based on an integral
over the emission line.

\subsection{Galaxy Morphology}\label{sub:morph}
\begin{figure}\epsscale{1}
%\plotone{f1new.jpg}
\plotone{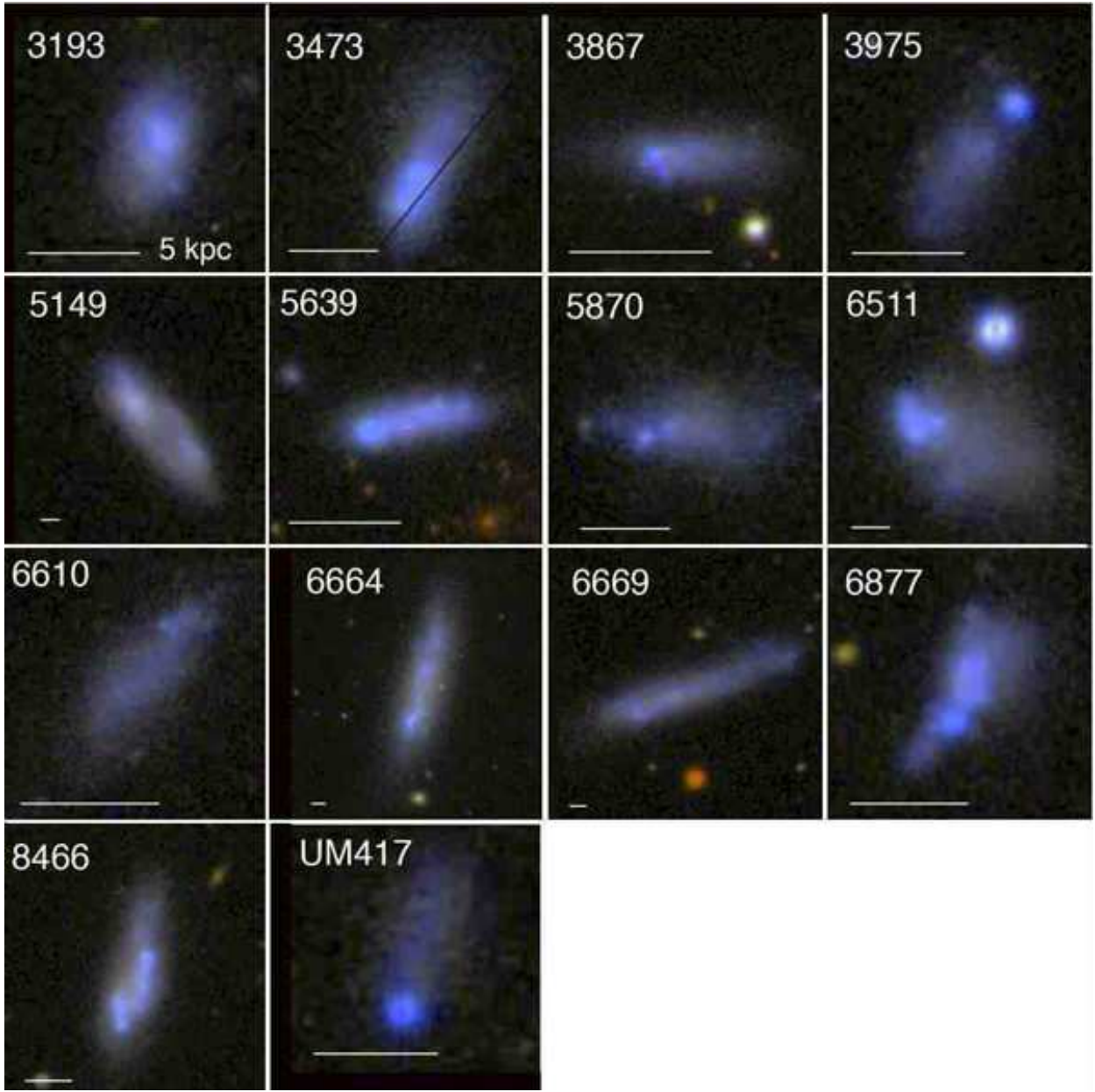}
\caption{Tadpole galaxies from the Kiso and UM samples. The white bar in each image
represents 5 kpc.}\label{tad}\end{figure}

\begin{figure}\epsscale{1}
%\plotone{f2new.jpg}
\plotone{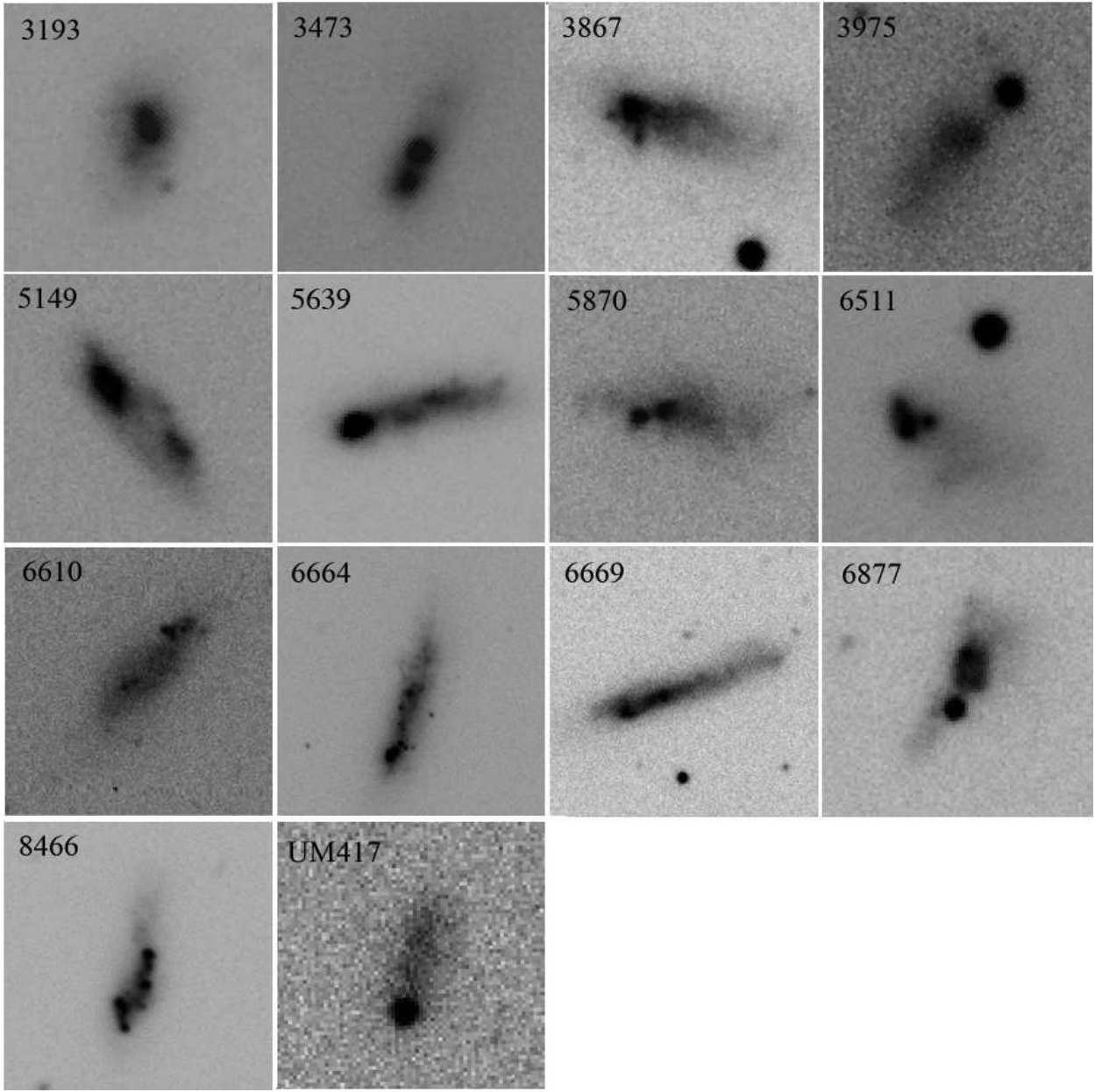}
\caption{Tadpole galaxies from the Kiso and UM samples, with the same order as Figure
\ref{tad}, shown in $g$-band grayscale for greater contrast. The individual
star-forming clumps are more obvious in this figure.}\label{gray}\end{figure}

The fourteen local tadpole galaxies in this study are shown in color in
Figure \ref{tad} and in grayscale in Figure \ref{gray}, and listed in Table
\ref{tab1}. The KUG numbers come from the Kiso2000 catalog \cite{nakajima}.
For convenience we also give a Kiso number, which is the catalog entry number
minus 1.

The tadpoles in Figure \ref{tad} are dominated by a large star-forming
region on one side, sometimes with sub-clumps, and by a diffuse tail, which
might also contain a few clumps. Some of the underlying galaxies are
elongated, such as Kiso 3473, 3867, 3975, 6664, 6669, and UM 417. Others are
rounder, such as Kiso 3193 and 6511. Galaxies 5149, 6664, and 6669 differ
from the others in being larger and more elongated, suggesting a highly
inclined disk. Others could be disks too, although some look too irregular in
the tails to be disks.

The smaller star-forming clumps are more easily seen in the grayscale images
than in the color images. Galaxies Kiso 3975 and UM 417
\citep{cairos1,cairos2} contain only one large round clump in the head that
is not obviously subdivided. Galaxies Kiso 3473 and 6877 have large head
regions with two sub-clumps in each. Kiso 5870 has a small star-forming
region in the head and other clumps elsewhere, including the tail. Kiso 6610,
6664, 6669, and 8466 have many small clumps over fairly large regions, while
Kiso 3867 and 6511 have many small clumps in their heads.

Five of these galaxies have been imaged in H$\alpha$ in previous studies:
Kiso 3473 and UM 417 \citep{paz},  Kiso 6511 and 6664 \citep{heller99}, and
Kiso 8466 \citep{gav}, which are all BCD galaxies. Their H$\alpha$ emission
stands out in the images just in the bright head regions and not in the
tails. \cite{heller00} studied lopsidedness in local dwarf Irregular galaxies
and found that random star formation could account for the asymmetry
distributions they observe in low surface brightness galaxies, but not in
BCDs, which have more centrally concentrated star-forming regions. Their BCDs
include three tadpoles, Kiso 6664 in our sample plus Mrk 5 and UM 133.

Although many star-forming galaxies at high redshift are large, the UDF
tadpole galaxies tend to be the smallest of the clumpy types, with diameters
ranging from 1 to 10 kpc and averaging about 4 kpc \citep{elm10}. The local
tadpoles selected for the present paper have larger average sizes than the
UDF tadpoles, with diameters ranging from 3 to 45 kpc based on their
isophotes, as described in Section \ref{sect:rad}.  Still, most of the
tadpoles are fairly small. Table \ref{tab1} indicates an average $R_{25}=6.7$
kpc and a median $R_{25}=3.2$ kpc, where $R_{25}$ is the radius out to a
surface brightness in $g$ of 25 mag arcsec$^{-2}$. Excluding the four largest tadpoles in our sample (radii greater than 10 kpc), the remaining ten tadpoles have average and median radii of 2.7 kpc, so their corresponding average diameters of 5.4 kpc are only slightly larger than the UDF tadpoles.

\subsection{Radial profiles}\label{sect:rad}
\label{sect:prof} The radial light distributions are complicated in
tadpole galaxies;  the galaxies are asymmetric and the profiles do not
have a continuous slope, so it is difficult to measure a S\'ersic index
in a meaningful way. Nevertheless, the S\'ersic index for UM 417 was
measured by \cite{amorin} to be 0.68$\pm$0.12. For comparison, S\'ersic
indices of high redshift clump galaxies are typically $\le$1 (e.g.,
Elmegreen et al. 2007). These rather flat disks are in contrast to the
tadpole BCDs Mrk 59 and Mrk 71, which \cite{noeske00} found to have
symmetric exponential profiles beneath the starbursts.

\begin{figure}\epsscale{1}
%\plotone{f3new.jpg}
\plotone{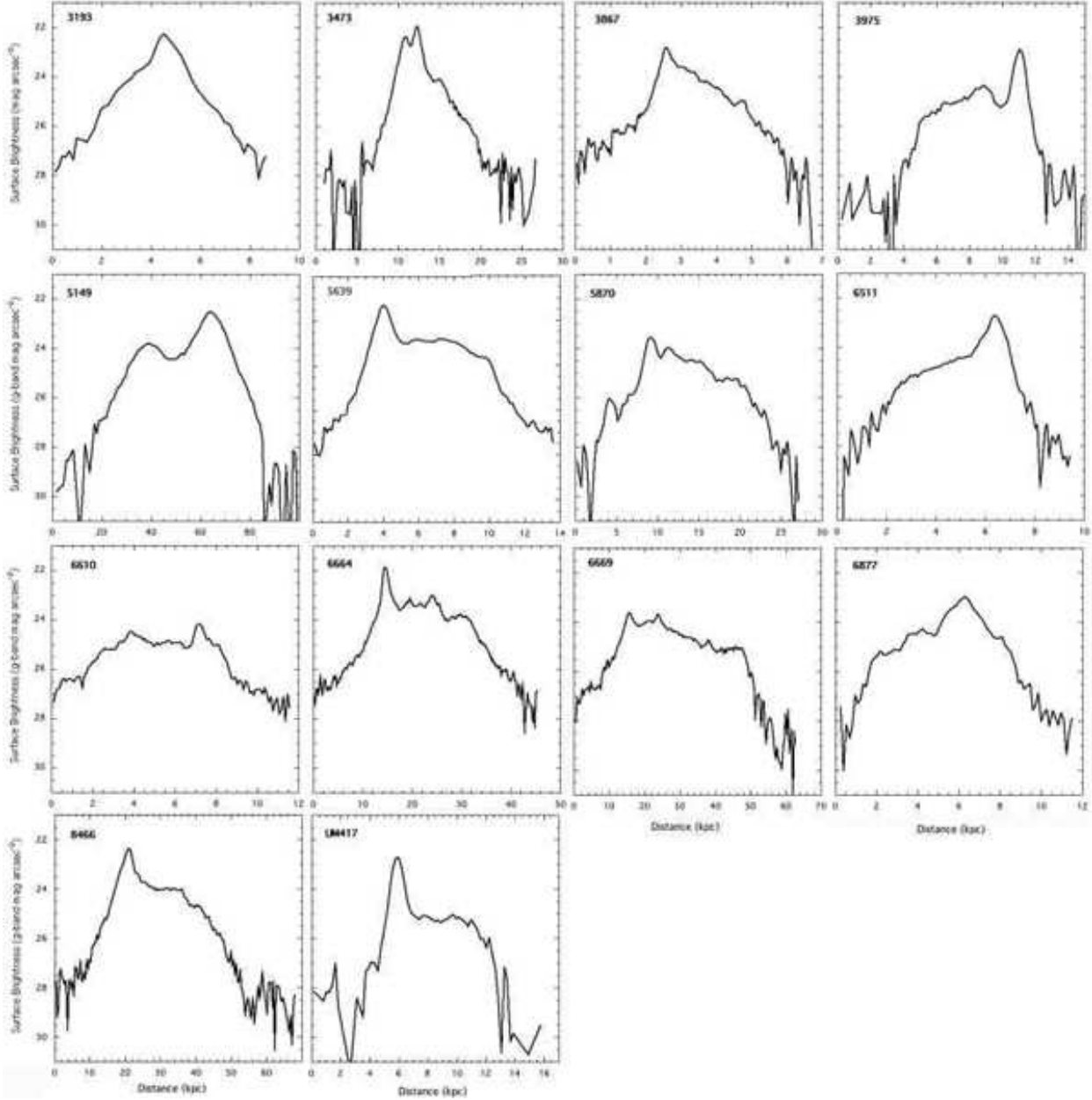}
 \caption{Radial profiles along the major axes of the tadpole
galaxies from the $g$-band filter; the other 4 filters are
similar. Exponential fits were made piece-wise for the left and
right outer parts and for the inner regions outside of the bulge.}
\label{tadrad}\end{figure}

For the local tadpoles in our sample, intensity profiles were made to examine
the light distributions in the outer parts of the galaxies. Cuts were taken
along the major axis, through the peak intensity of the head. The intensity
cuts were five pixels wide, corresponding to 5.75$^{\prime\prime}$, to reduce
the noise.  The major axes of the galaxies were determined by eye from
contour plots and grayscale $g$-band images.  The $g$-band surface brightness
profiles are shown in Figure \ref{tadrad} (other filters have similar
profiles).  The head is clearly an excess intensity on one side. For all but
Kiso 3193, the galaxy intensity does not decrease exponentially and
symmetrically away from a central position defined by the outer isophotal
contours. The profiles are usually flat or have a shallow exponential
decrease on one side of the head in the central third of the galaxy. They
have sharper drop-offs beyond $\sim25.5$ mag arcsec$^{-2}$ and can be traced
to $\sim28$ mag arcsec$^{-2}$ in all SDSS filters. The radii $R_{25}$ of the
galaxies, out to a surface brightness of 25 mag arcsec$^{-2}$ in $g$ band,
are listed Table \ref{tab1}. These are determined from the diameter of the
galaxy to this limiting surface brightness based on the radial profiles shown
in the figure; that is, they are the half-length of the major axis contours
out to a surface brightness of 25 mag arcsec$^{-2}$ in $g$ band. The surface
brightness of the sky background is typically 28 to 28.5 mag arcsec$^{-2}$ in
$g$ band. The sharper drop-offs at the edge are approximately exponential.

We fit piece-wise exponentials to the intensity profiles in Figure
\ref{tadrad}.
%and to the same major axis strips in other passbands.
Three pieces were considered: the relatively flat inner part of the galaxy
outside the head, and the two relatively steep outer parts beyond $\sim25.5$
mag arcsec$^{-2}$ in the left and right extremes of the radial profiles in
Figure \ref{tadrad}. Table \ref{tab2} lists the scale lengths $r_s$, the
scale lengths relative to $R_{25}$, and the number of scale lengths in the
fitted regions. The inner parts of the disks have large $r_s$/$R_{25}$,
reflecting the flatness of the diffuse distributions seen in Figure
\ref{tadrad}. The extreme cases are represented by Kiso 8466 and UM 417, with
length scales in excess of 100~kpc. The inner disk scale length in Kiso 5149
was not measured because there are two bright clumps in the profile there.

Table \ref{tab2} shows that the left and right outer scale lengths are often
similar to each other and much smaller than the inner scale lengths. This
similarity between the far-outer scale lengths suggests there is a faint
underlying symmetric structure, like a disk. In Kiso 3193 and 3867, the
slopes are about the same in the inner and outer parts. The profile of UM 417
was measured by \cite{cairos1} as part of a photometric study of the light
distributions in 28 BCDs; its scale length in that paper matches our value
for the left side of the galaxy.

\subsection{Photometry}\label{sub-phot}
Photometric procedures for measuring the local tadpoles were the same as
those used in our previous studies of UDF high redshift galaxies. Heads were
measured with the task {\it imstat} by defining a rectangle around the
largest star-forming complex in each galaxy, which generally contained more
than one resolved sub-clump. This task gives the average surface brightness in a pixel, which is multiplied by the total number of pixels in a clump to determine the magnitude in a given filter.  For comparison we also measured the head magnitudes with
circular apertures in the {\it apphot} task.
%Regions are often elongated rather than circular, so a rectangle allows a more precise fit.
These two measurements gave the same results, differing by less than the
measuring error of 0.1 mag. The rectangular results are presented in the
figures since this is the same method used for high redshift galaxies.
Individual sub-clumps in the heads and small clumps in the tails were also
measured with {\it imstat}. The average intensity of the brightest clump in
each galaxy is about 10 times the average intensity of its tail, and about
50$\sigma$ above the tail intensity variation.

Clump emission was determined in two ways. In one method, just the sky was subtracted. In the second method, we assumed the clumps are sitting in disks, and subtracted a smooth component
of the galaxy from the clump to obtain the emission from
the star-forming part alone.
%This smooth component was difficult to estimate because the head is not superposed on a symmetric exponential disk.
For this subtraction, we used the average tail surface brightness from a
section between clumps inside the tail, measured in a rectangular region with
{\it imstat}. This is appropriate if the head or clump is superposed on a
stellar disk which extends into the tail. Subtracting the underlying
component usually makes the clumps bluer and therefore younger and less
massive than with sky subtraction alone. The quantitative differences are
discussed in Section \ref{pop} below. The clump and tail surface brightnesses
were also used to determine surface densities, discussed in Section
\ref{surf}.
%The total mass surface density in the head region equals the sum of the tail surface density and the surface density
%of the star-forming part of the head.

In addition, the total magnitudes of the tails were determined from a
rectangular outline enclosing the whole tail, minus the sky emission. These
total magnitudes were used to determine tail ages and masses, as described
below.

\subsection{Population synthesis fits}\label{pop}
Masses, mass surface densities, ages, and extinctions were estimated from our
photometry by fitting the data to population synthesis models from
\cite{bruzual}. To determine the model colors, we integrated over the low
resolution spectra in \cite{bruzual}, weighted by the Gunn filter functions,
and assumed a constant star formation rate back to some starting time, which
is identified with the region age.  The Chabrier IMF was also assumed. A
range of extinctions was used for the models, following the wavelength
dependencies in \cite{calz00} and \cite{leith02}.  We used a metallicity of
0.4 times solar since we expect these small galaxies to have slightly
sub-solar metallicity. This metallicity matches what is observed based on
emission line flux ratios in the BCD galaxy Mrk 1418 \citep{cairos}. We also
ran models with decaying star formation rates and with metallicities of 0.02 solar, 0.2 solar, and solar metallicity
for comparison. The final values for mass, age, and extinction were
determined as the weighted average of the different models in our standard case (i.e., constant star formation rate and metallicity of 0.4 solar), using a weighting
function $e^{-0.5\chi^2}$ for $\chi^2$ equal to the quadratic sum of the
differences between the model and observed colors, normalized to the relative
dispersions for each color from the tail observations. The weighted averages
were done only for the models with the lowest rms deviations from the
observations, as determined by binning the rms deviations in intervals of
0.02.  The mass follows from the observed $g$ magnitude in comparison to the
model. The surface densities in the clumps are taken to be the masses divided
by the projected areas.

To determine the effects of background subtraction and metallicity on ages
and masses, we make the following comparisons. First consider the results for
solar metallicity. When only the sky brightness is subtracted, the average
$\log$(age) of the heads is $8.3\pm1.3$ (for age in yr). When the sky is
subtracted along with an underlying component equivalent to the average tail
brightness, the heads are younger on average, $\log({\rm age})=7.5\pm1.3$.
(The rms errors are the quadratic sum of the rms errors for each galaxy and
the rms of the average of the different galaxies.)  For 0.4 solar
metallicity, these values of $\log$(age) are $8.4\pm0.9$ and $7.9\pm0.8$,
respectively. Lower metallicity increases the calculated age because low
metal stars are intrinsically bluer than high metal stars, so their ages have
to be larger to get the same observed colors. The head ages with only the sky
subtracted are older than the head ages with both the sky and an underlying
red emission subtracted, because the head still contains old and red emission
when only the sky is subtracted.

The average $\log$(mass) of the heads is $7.1\pm1.0$ (for mass in $M_\odot$) when
only the sky brightness is subtracted for solar metallicity. The heads are
less massive, $\log({\rm age})=6.7\pm1.0$, when the sky and underlying red
component are subtracted.  For 0.4 solar metallicity, these log(mass) values
are $7.0\pm1.0$ and $6.6\pm1.0$, respectively. Thus, decreasing metallicity
hardly affects the clump masses. Removal of a red underlying component
decreases the clump masses by an average factor of $\sim3$.

For whole tails with sky subtracted, the average log(age) is much older than the heads, $9.4\pm0.4$,
$9.5\pm0.3$, $9.4\pm0.3$, and $9.4\pm0.3$ for metallicities of $0.02$, $0.2$,
$0.4$ and solar, respectively, not counting the galaxy UM 417. For UM 417,
the tail log(age) is much less than the others, $\sim7.1$, and it has a much
larger uncertainty than the others, $\sim1.5$, for the highest three
metallicities. At 0.02 solar for UM 417, the log(age) is like the others,
$9.3\pm0.4$. Aside from UM 417, the metallicity does not affect the tail age
much.

Also for the whole tails with sky subtracted, the
average log(mass) among the sample is $7.8\pm0.9$, $7.7\pm0.9$, $7.7\pm0.9$,
and $7.8\pm0.9$ for the four metallicities, respectively, whether or not UM 417 is included.
Thus, tail masses do not have a strong systematic dependence on metallicity either.

The average equivalent $V$-band extinctions in the heads with the underlying disk
subtracted are $0.9\pm0.7$, $0.6\pm0.6$, $0.5\pm0.5$, and $0.7\pm0.7$ for
metallicities of $0.02$, $0.2$, $0.4$ and solar, respectively. For the tails,
the extinctions are $0.8\pm0.5$, $0.4\pm0.4$, $0.5\pm0.4$, and $0.5\pm0.4$,
respectively.

Models with decaying star formation rates have ages that are slightly less than the models with constant rates. This is because a decaying rate has proportionately less star formation today than a constant rate, and therefore needs to start more recently to end up with the same observed colors. For models of the head with both sky and an underlying emission subtracted, and for 0.4 solar metallicity, the log of the age decreases by 0.02, 0.06, and 0.22 as the decay time decreases from infinity (i.e., constant star formation rate) to $3\times10^9$ yr, $1\times10^9$ yr, and $3\times10^8$ yr, respectively. For the whole tails with sky subtracted, the log of the age decreases by 0.20, 0.18, and 0.62, respectively. The masses are relatively unchanged along this sequence, with the log of the mass changing by less than 0.01 for the heads, and by 0.1 to 0.2 for the tails.  These examples indicate that the ages and masses we discuss below are not very sensitive to the star formation history. 

In general, the age of an extended region is difficult to determine accurately from the
small number of broad-band magnitudes used here.  The mass estimates are more
reliable because extinction and age compensate for each other in the
conversion from color to mass.  In what follows, we use head and clump values
with an underlying sky and disk subtracted because we are interested in the
star formation components. We also use 0.4 solar metallicity for the
discussion below and for the ages and masses in Table \ref{tab3}, because
many of the tadpoles are small galaxies. The star formation rate is assumed to be constant to minimize the number of free parameters.

\section{Results}
\label{results}
\subsection{Head and Tail Masses and Surface Densities}
\label{surf} 
\begin{figure}\epsscale{1}
%\plotone{f4new.jpg}
\plotone{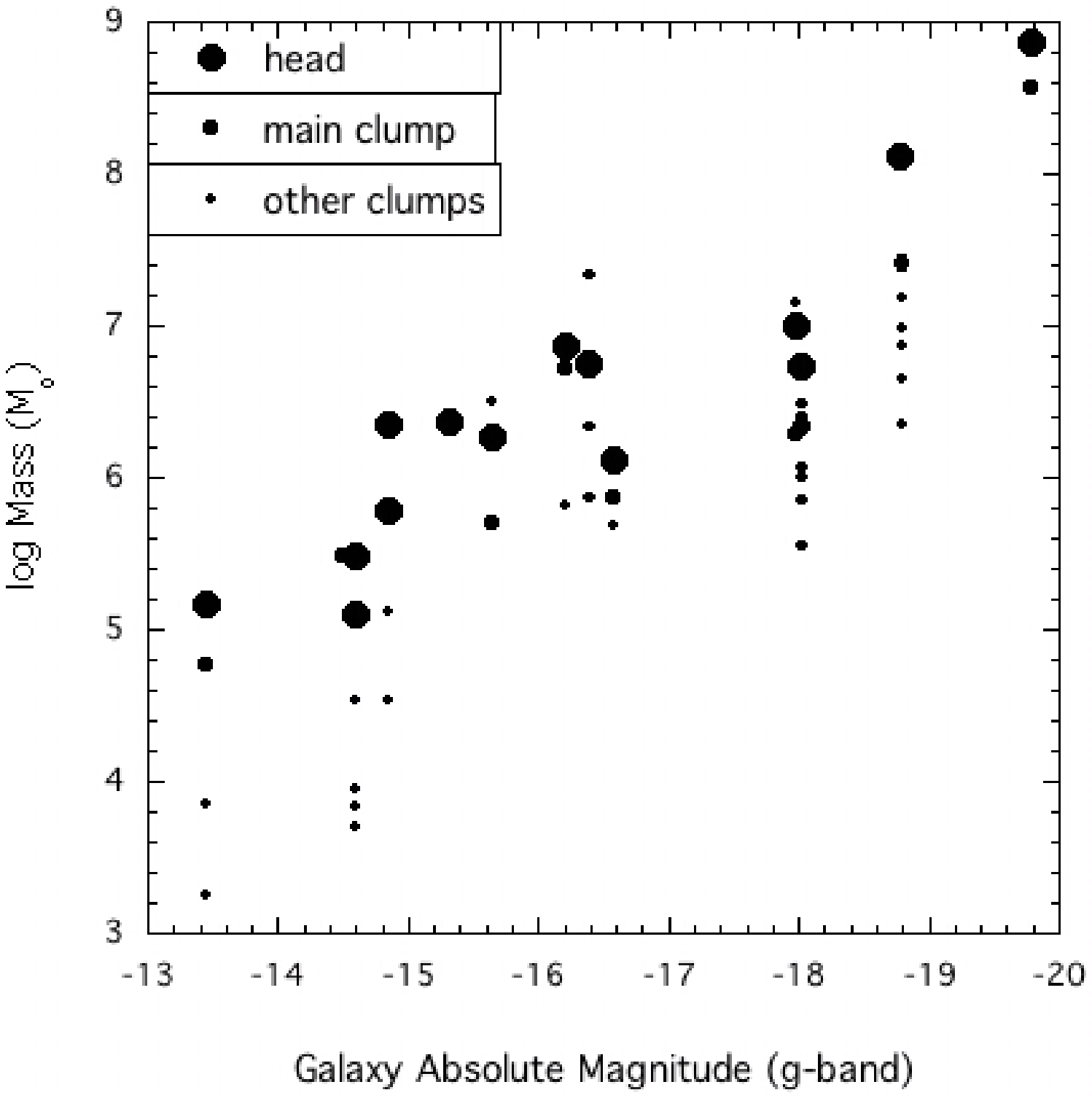}
\caption{Stellar masses of the heads (large dots), the biggest clump in each
galaxy (intermediate dots),
and all of the other clumps (small dots), for local tadpole galaxies in the
Kiso and UM samples, are plotted as a function of the galaxy absolute $g$-band
magnitude. The masses were derived from photometric results  with underlying disk subtraction and assuming 0.4 metallicity. The small clump with log(mass)=7.4 at magnitude -16.4 has a large mass from the model fits because it is older than the other clumps; it shows up as a reddish clump to the right of the head in Kiso 5639.}\label{massall}\end{figure}

Stellar masses of the tadpole heads are listed in Table
\ref{tab3}. They range from a few $\times10^5\;M_\odot$ to a few
$\times10^8\;M_{\odot}$. Some of the heads resolve into sub-clumps, as in
Kiso 6511. Figure \ref{massall} shows, as a function of galaxy $g$-band
absolute magnitude, the masses of the heads, the masses of the brightest
single clumps in the heads, and the masses of all the other clumps. The most
massive clumps are larger in more luminous galaxies. This trend is partly a
selection effect because the galaxies are chosen to have similar morphologies
-- all of the features in the morphology are larger for larger galaxies.

\begin{figure}\epsscale{1}
%\plotone{f5new.jpg}
\plotone{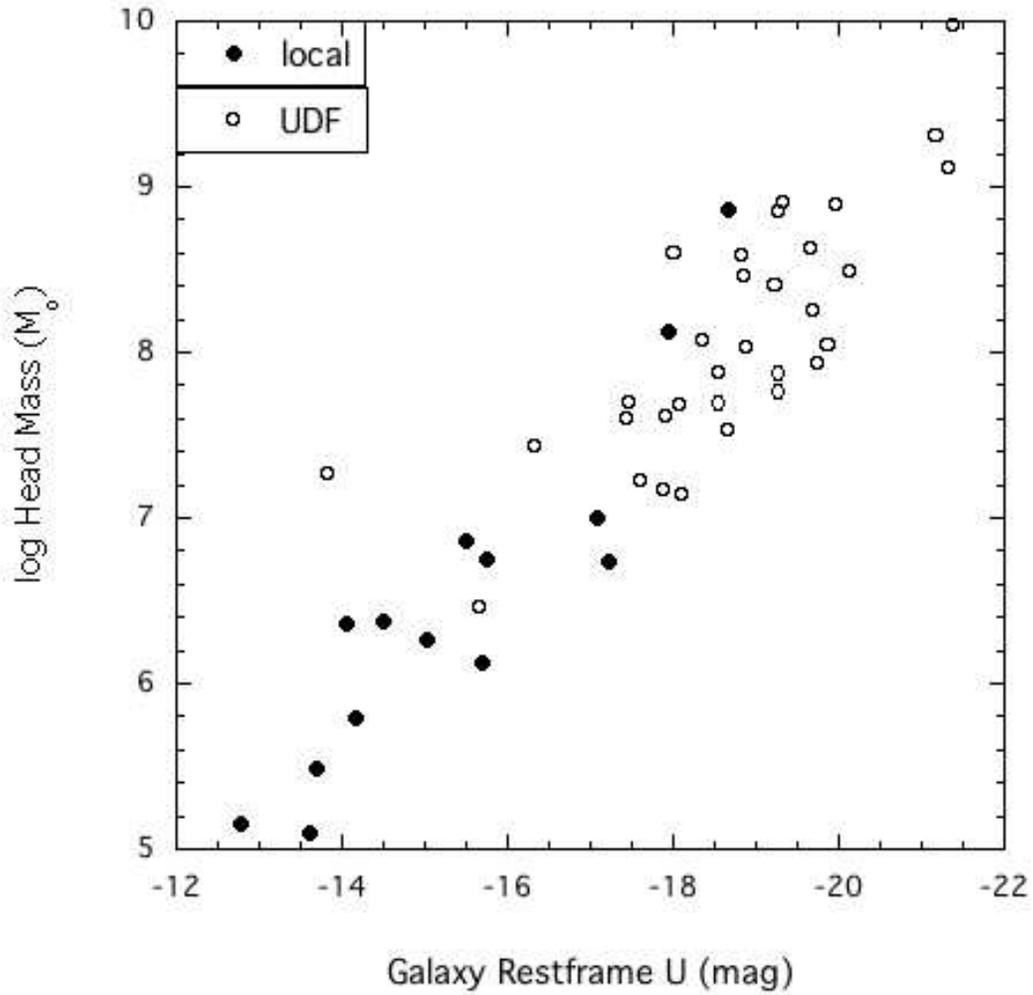}
\caption{Head masses for the local tadpole galaxies, shown as black dots, and for the
high redshift UDF tadpoles, shown as open circles, plotted as a function of galaxy
restframe U magnitude. The head masses are approximately proportional to galaxy luminosity.}\label{mass}\end{figure}

Figure \ref{mass} shows the log of the head masses for both local and high
redshift tadpoles as a function of the restframe $U$ magnitude. The $U$ magnitude
is determined for the high redshift galaxies from an interpolation of the
observed magnitudes in ACS or NICMOS filters, depending on the redshift
\citep[which is from][]{rafel09}. The $U$ magnitude for a local galaxy is
assumed to be the absolute magnitude determined from the $u$ image, using the
distances listed in Table \ref{tab1}. The heads in the local tadpoles are, in
general, less massive than the heads in high redshift tadpoles \citep{elm10}.
In all cases, the head mass is approximately proportional to the galaxy
luminosity. The brightest local tadpoles overlap the range for the high
redshift tadpoles, and in those cases the head masses are comparable locally
and at high $z$. The two brightest local tadpoles are Kiso 5149 and Kiso
8466. Kiso 5149 is one of the largest galaxies in the local sample, with a
radius of 22.9 kpc. This galaxy looks different from the others in being
redder and older. Kiso 8466 has a massive head with many smaller clumps over
a large region, which at high $z$ would appear as one or perhaps two large
clumps. The scaling in Figure \ref{mass} is partly a selection effect based
on the similar morphologies of low and high redshift tadpoles.

\begin{figure}\epsscale{0.8}
%\plotone{f6new.jpg}
\plotone{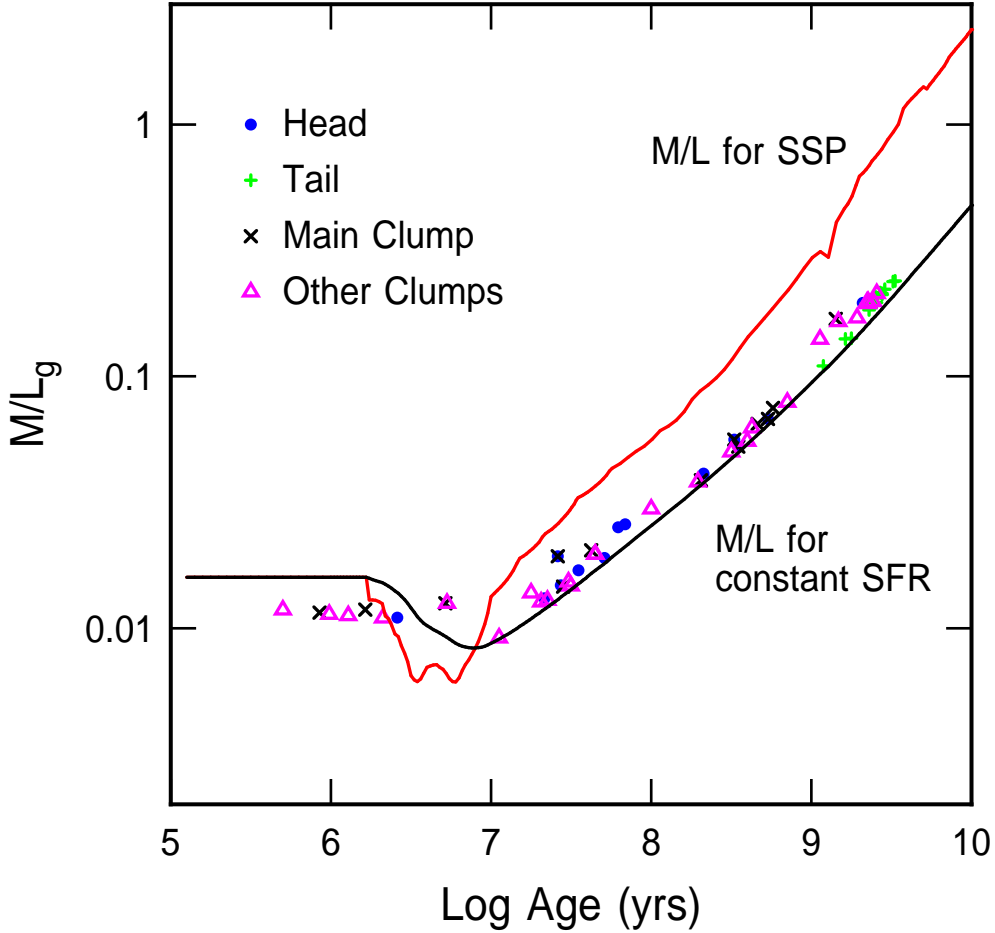}
\caption{Mass-to-Light ratios for tadpole components versus the ages, as determined
from population synthesis fits. The top curve shows the $M/L_{\rm g}$ ratio
for single stellar population models (SSP) with a burst of star formation at the
age on the abscissa. The bottom curve shows the
$M/L_{\rm g}$ ratios for SSP models integrated over time from the present
time to the age on the abscissa, as representative of $M/L_{\rm g}$ for
continuous star formation. The observational fits use the same continuous
star formation model as the bottom line. The observations deviate slightly
from the bottom curve because the observations are best-fit averages to a
wide range of models. Five points have been omitted because the rms uncertainties in their log(age) determinations exceed 1. }\label{ml}\end{figure}

The ratio of mass to $g$-band luminosity, $M/L_{\rm g}$, is shown as a
function of fitted age for the heads, tails, main clumps, and other clumps,
in Figure \ref{ml}. The four galaxy components discussed in Figure
\ref{massall} are shown by different symbols. $M/L_{\rm g}$ increases
smoothly with age, as expected for population synthesis models. The top curve
is the $M/L_{\rm g}$ ratio for single stellar population (SSP) models with a single burst of star
formation at the time on the abscissa, made using tables in \cite{bruzual}.
The bottom curve is for an integral of the SSP model over time from the
present back to the time on the abscissa. $M/L_{\rm g}$ from the fits of the
model to the observed colors matches the integrated case, as it should,
because we assume a constant star formation rate over the age of the
region. There is a clear separation between the tails, which have an average
log(age) of $9.4\pm0.3$ for ages in years and average  $M/L_{\rm
g}\sim0.19\pm0.11$, and the heads, which have an average log(age) of
$7.9\pm0.8$ and $M/L_{\rm g}\sim0.043\pm0.061$.

\begin{figure}\epsscale{1}
%\plotone{f7new.jpg}
\plotone{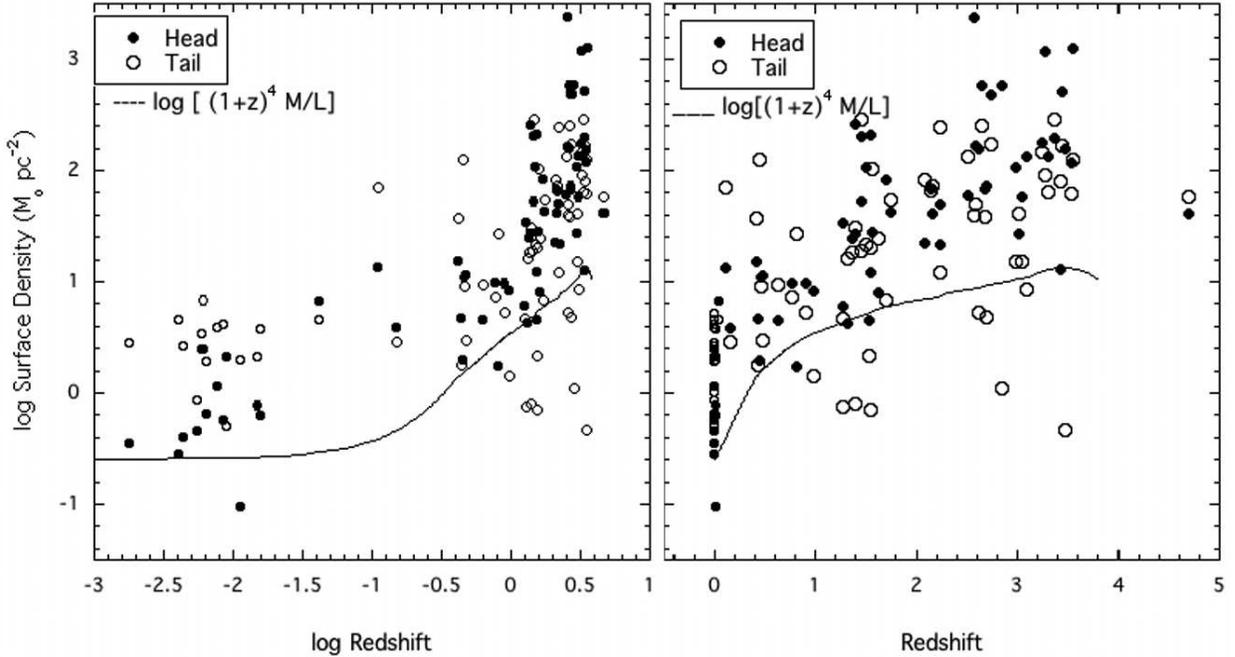}
\caption{Surface densities of tails and heads for tadpole galaxies from the local and
UDF samples. Head surface densities have a background disk subtracted. (left)
The abscissa has log (redshift) to emphasize the local
galaxies. (right) The abscissa is linear in
redshift. The solid lines show the trend of the observable surface
density for a surface brightness of 1 $L_{\odot}$ pc$^{-2}$ produced by
cosmological dimming, bandshifting, intervening hydrogen absorption, and an M/L ratio that varies with time at constant star formation rate since $z$=4.}
\label{dens}\end{figure}

The head surface densities, $\Sigma_{\rm head}$, in units of
$M_{\odot}$ pc$^{-2}$, are listed in Table \ref{tab3} and shown as filled
circles in Figure \ref{dens} as a function of redshift.  The surface
densities range from $\sim0.09\;M_{\odot}$ pc$^{-2}$ to $\sim6.6\; M_{\odot}$
pc$^{-2}$ for the local tadpoles, with a mean value of $1.4\pm1.6\; M_{\odot}$
pc$^{-2}$. The UDF tadpoles are shown as open circles for comparison; their
$\Sigma_{\rm head}$ values range from $\sim2\; M_{\odot}$ pc$^{-2}$ to
$3000\; M_{\odot}$ pc$^{-2}$. Below redshift $z=1$, the UDF tadpoles have
$\Sigma_{\rm head}\sim2\; M_{\odot}$ pc$^{-2}$ to $20\; M_{\odot}$ pc$^{-2}$,
only slightly higher than the local tadpoles.  The left-hand panel shows log
redshift on the abscissa to emphasize the local tadpoles, while the
right-hand panel shows linear redshift to emphasize the UDF tadpoles.  The
average $\Sigma_{\rm head}$ steadily increases with redshift. This increase
is consistent with the increasing star formation rate in the universe, which
peaks between $\sim z=2-4$ \citep{madau,bouwens11}. This suggests that
tadpoles are more gas-rich at earlier times, like other galaxies.

The general increase of surface density with redshift is partly a selection
effect. Surface density is measured from surface brightness, and higher
redshift galaxies need higher intrinsic surface brightnesses in order to be
seen. This is primarily because of cosmological surface brightness dimming. Our
previous studies \citep{elm09} showed a systematic increase in intrinsic
surface brightness as $(1+z)^4$, which is the expected selection effect.
Surface mass density does not increase as rapidly because the disks get
younger and their $M/L$ ratios decrease with $z$.

The curve in Figure \ref{dens} plots log $(1+z)^4M/L_{\rm g,rest}$ with
redshift, obtained from the spectral models in \cite{bruzual} with the
$g$-band SDSS filter band-shifted into the restframe $g$-band as redshift
increases. This bandshifting was done by redshifting the model galaxy spectrum while keeping the wavelength distribution function of the filter constant.  Intervening hydrogen absorption is included \citep{madaup}. The plotted 
function shows the redshift trend for the threshold mass surface density at a
fixed limit of observed surface brightness, equal to 1 $L_{\odot}$ pc$^{-2}$. The galaxies tend to lie above this line because their surface brightnesses are larger than this fiducial value.
The model assumes that star formation starts at $z=4$ and is continuous until
the galaxy is observed at some lower redshift $z$. The initial increase in
$(1+z)^4M/L_{\rm g,rest}$ with increasing $z$ is mostly from $(1+z)^4$. There
is a slight bump above this $(1+z)^4$ trend at $\log z\sim-0.38$ because
$M/L_{\rm g,rest}$ has a peak there, as follows. At lower $z$, $M/L_{\rm
g,rest}$ increases with increasing $z$ because the age of the star-forming
region decreases from 4.3 Gyr (at $\log z=-3$) to 3.7 Gyr (at $\log z=-0.38$)
and $L_{\rm g,rest}$ drops faster than $M$ as the restframe passband moves
from green into the aged blue and ultraviolet parts. At $\log z$ higher than
$-0.38$, $M/L_{\rm g,rest}$ decreases with increasing $z$ because the stellar
population gets younger and bluer with decreasing age, $M$ continues to
decrease steadily, and $L_{\rm g,rest}$ increases as the restframe passband
moves into the ultraviolet where the young stars are brightest.

Tail masses and surface densities are listed in Table \ref{tab3}. Tail masses
are on average a factor of 12 times larger than head masses because the tails
are larger in area and surface density than the heads. The tail stellar
surface densities are shown as open circles in Figure \ref{dens}. The larger
tail surface densities follow from the older tail ages and the higher
$M/L_{\rm g}$ (Fig. \ref{ml}) for the tail compared to the head.  Some of the
tails have clumps that are included in the category of ``other clumps" in
Figure \ref{massall}.

The tails have a mean mass surface density of $3.0\pm1.4\;
M_{\odot}$ pc$^{-2}$ seen in projection (not corrected for inclination). The
mean surface density for the high $z$ tails is about the same, $3.0\pm3.3$
$M_{\odot}$ pc$^{-2}$, but there are many with higher surface densities at
high $z$. These high-$z$ surface densities include corrections for surface
brightness dimming and intervening hydrogen absorption.

The ages of the tails differ at low and high redshift. Local tadpole tails
have ages of 1-3 Gyr (the $\log$ of the average age in Table \ref{tab3}, in
years, is 9.2$\pm$0.8, or $9.4\pm0.3$ without UM 417), while high redshift
tails have ages of $\sim0.1$ Gyr.  The high-$z$ systems are visible because
of their greater luminosity per unit mass.

The heads have about the same ages for local and high redshift tadpoles,
$0.1-0.2$ Gyr. These ages are reasonable for giant star-forming regions. The
head and tail ages are similar at high redshift but different at low redshift
because the UDF galaxies are younger overall than the local galaxies.

For high $z$ tadpoles, the average of the log of the ratio of tail surface
density to head surface density is $-0.34\pm0.92$. For local tadpoles, the
average of the log of the ratio is much higher, $0.53\pm0.52$. The increase
in tail-to-head surface density ratio with decreasing redshift indicates that
the head star formation rate per unit area gets smaller compared to the tail
surface density as the redshift decreases. This is the expected trend if
tadpoles are normal galaxies with a gas fraction that decreases and a stellar
mass that increases over time. The same is true for star formation in other
local galaxies: star complexes today contain only a few hundredths of the
surface density of their surrounding disk, whereas high redshift star-forming
clumps contain a comparable surface density or more \citep{elm09}.

The surface densities of both the star-forming heads and the tails of local
tadpole galaxies are much smaller than the stellar surface densities of local
spirals, which are $\sim100\;M_\odot$ pc$^{-2}$. In the solar neighborhood,
it is $\sim70\;M_\odot$ pc$^{-2}$ \citep{freeman87} for the thin disk
component. Low surface brightness galaxies and local dwarf irregulars have
surface densities as low as those observed for local tadpole tails.
\cite{zhang12} find $\Sigma\sim1-10\;M_\odot$ pc$^{-2}$ for 34 dwarf
irregulars.  This similarity in surface density between local tadpoles and dwarf irregulars is consistent with the small
sizes of the local tadpoles, as indicated by $R_{\rm 25}$ in Table
\ref{tab1}.

\subsection{Star formation rates}
\label{sub-sfr} 
\begin{figure}\epsscale{1}
%\plotone{f8new.jpg}
\plotone{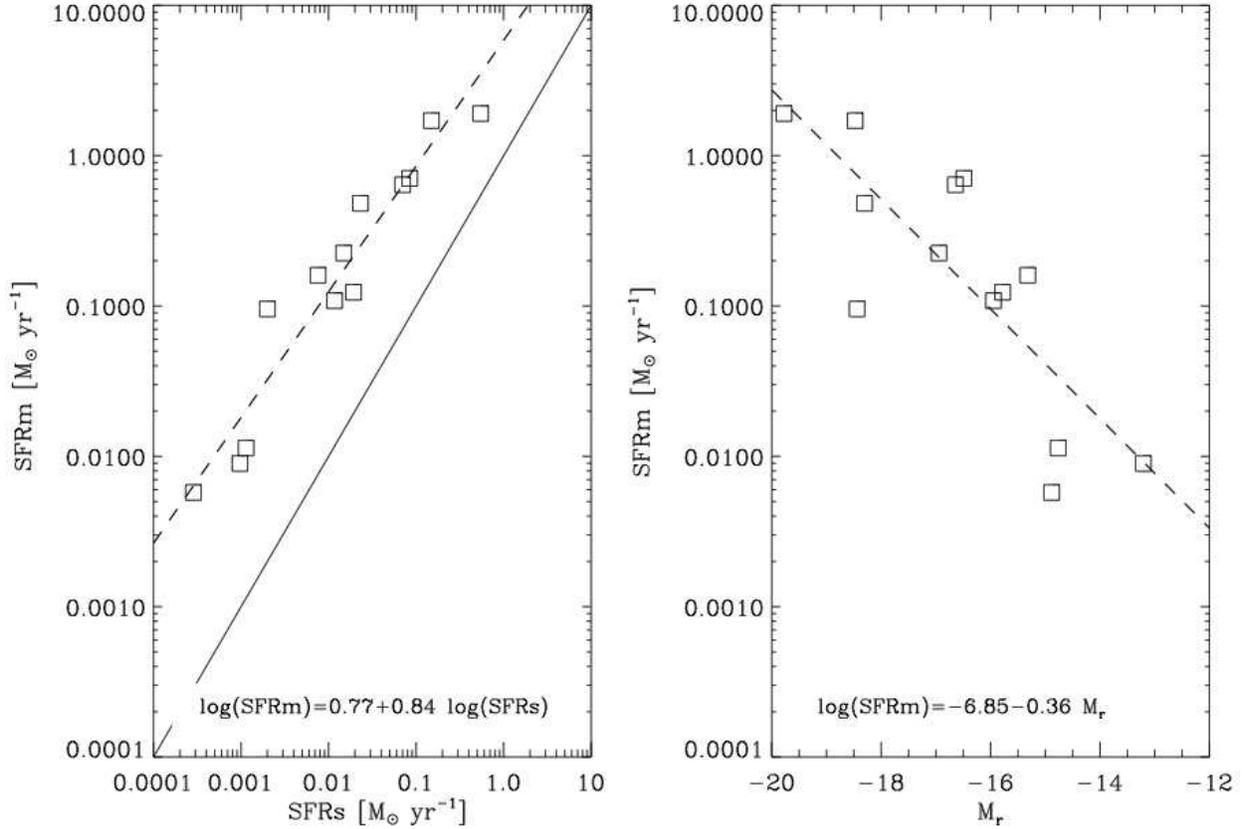}
\caption{(left) Comparison between the star formation rate in the central
3$^{\prime\prime}$
averaged by the SDSS spectrum (SFRs), and that for the whole galaxy obtained
by scaling the observed spectrum with the spatially integrated $r$-band flux
(SFRm).  The diagonal solid line indicates where the two star formation rates
are equal. (right) Star formation rates for the whole galaxies as a function
of the absolute $r$ magnitude from SDSS. The dashed lines show linear fits to
the scatter plots, with the corresponding  equations given as
insets.\label{cassfr}}\end{figure}

The average head and tail star formation rates (SFR) were
calculated from the ratios of the derived masses to the ages, and are listed
in Table \ref{tab4}. They represent an average over the lifetime of the
dominant stellar population. They differ from the {\em current} star
formation rates, which were estimated from the H$\alpha$ fluxes in the SDSS
spectra (available for all but UM~417). Star formation rates were calculated
in two ways from the spectrum: one from the $3^{\prime\prime}$-diameter area
of the SDSS fiber using the measured H$\alpha$ flux in the fiber, which was
usually centered on the brightest part of the galaxy, and the other from the
product of the H$\alpha$ equivalent width in the fiber spectrum (i.e., the
ratio of H$\alpha$ flux to $r$-band flux) and the integrated $r$-band
luminosity of the whole galaxy, using the conversion by \cite{ken}. These two
rates are also listed in Table \ref{tab4}. The first case corresponds to the
SFR in the $3^{\prime\prime}$ region covered by the SDSS spectrum, so it
usually underestimates the total galactic SFR. The second case assumes all of
the galaxy is forming stars at the relative rate of the fiber part, so it
represents what is likely an upper limit.  Figure \ref{cassfr} shows the two
H$\alpha$-based rates. The left side has a comparison of the two values,
showing that the value from the extrapolation to the whole galaxy is about
ten times larger than the value in the fiber only. The right side shows the
star formation rate extrapolated to the whole galaxy versus galaxy $r$-band
absolute magnitude M$_r$. The equations in the figures show the linear fits
to the data.

Previous studies determined star formation rates for some of the tadpoles.
For one of our galaxies, Kiso 3473 = Mrk 1416, \cite{zhao} found a value of
log SFR $= -1.37$ (SFR in $M_\odot$ yr$^{-1}$) using H$\alpha$ flux from an
SDSS spectrum. This value differs from our value of $-1.47$ in Table
\ref{tab4} because we used an SDSS spectrum from a different position in this
galaxy. This galaxy has two object identification numbers in SDSS, one for
each clump. When we convert the H$\alpha$ flux for the Zhao et al. position,
we confirm their value. H$\alpha$ luminosities were also measured for some of
our tadpoles in previous studies. For these, we can derive a SFR for the
whole galaxy by using the \cite{ken} conversion and applying the correction
for a metallicity of 0.4 solar (i.e., $Z=0.008$) as in \cite{hunter}: SFR
($M_{\odot}$ yr$^{-1}$) = $6.9\times10^{-42}$ $L_{H\alpha}$ (erg s$^{-1}$).
For Kiso 3473, an observed log H$\alpha$ luminosity $= 40.42$ \citep[units of
erg s$^{-1}$;][]{paz} gives log SFR $= -0.74$, and for UM 417,  log H$\alpha$
= 39.70 gives log SFR$= -1.46$. For Kiso 8466, log H$\alpha$ = 40.64
\citep{gav} gives log SFR $=-0.52$. \cite{heller99} give log SFR $= -1.43$
for Kiso 6511, and $-1.06$ for Kiso 6664. These values are in between the
SFRs we derive for the heads and those extrapolated for the whole galaxies.
%NOTE: Above, we use Kennicutt without metallicity conversion - be consistent. Will ref complain about saying the Halpha rates are higher since they include the whole galaxy - doesn't that mean some of it is from the tail, which we argued was not star-formng?

\begin{figure}\epsscale{1}
%\plotone{f9new.jpg}
\plotone{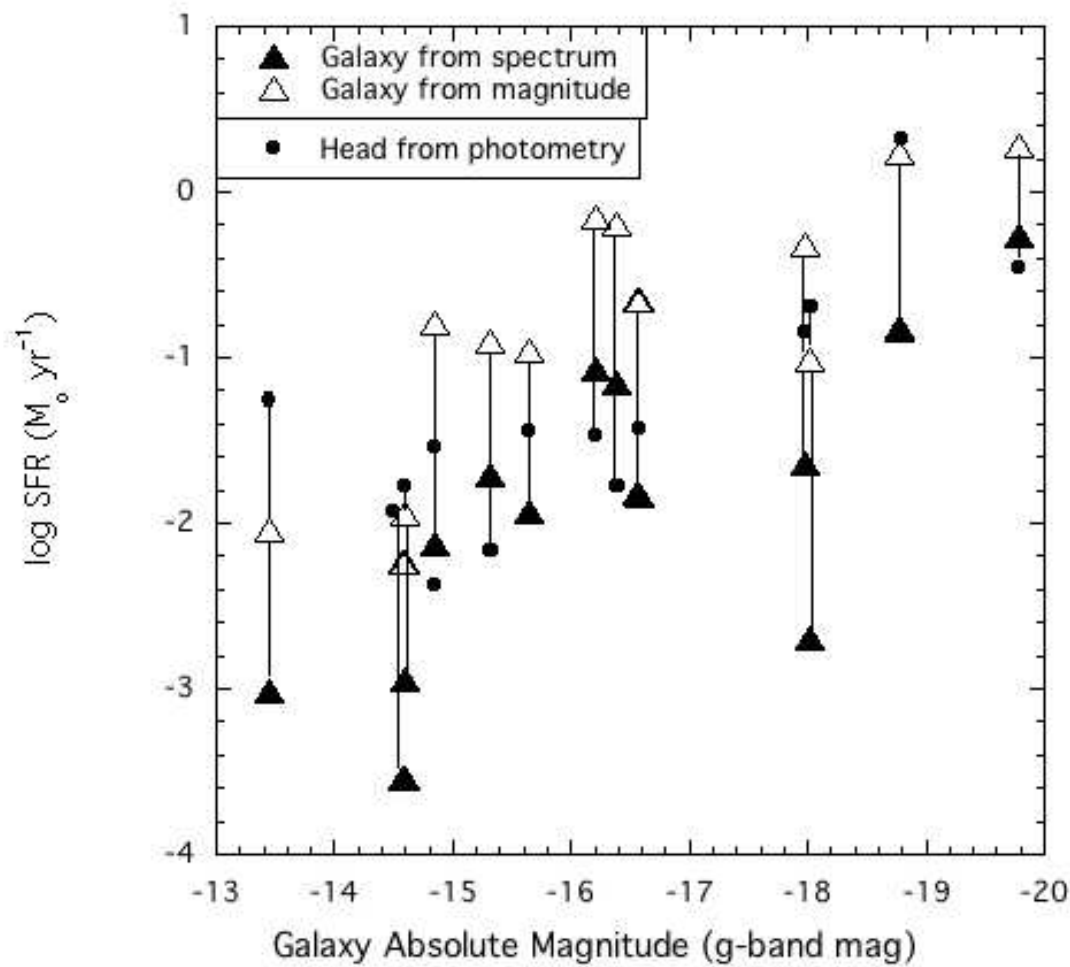}
\caption{Comparison between the three star formation rates estimated in the paper. The
{\em current} star formation rates inferred from H$\alpha$ fluxes are shown
as triangles: the filled triangles are for the star formation rates averaged
over the 3$^{\prime\prime}$ SDSS fiber, which represent a lower limit for the
galaxy, and the open triangles are for an extrapolation of these rates to the
whole galaxy based on the H$\alpha$ equivalent width scaled to the the
spatially integrated $r$-band magnitude. Filled circles are the ratio of the
star formation mass to the age of the heads, using broad-band colors; they
represent an average over the lifetime of the young stellar population. Thus
each galaxy has 3 points, all at the same magnitude. The galaxy UM 417 has
only one point because there is no spectral information. The abscissa is the
absolute $g-$band magnitude. }\label{sfr}\end{figure}

Figure \ref{sfr} shows the star formation rate for local tadpoles as a
function of the galaxy absolute $g$ magnitude. The three rates discussed
above are plotted: the head region from photometry, the fiber region from
H$\alpha$ flux in the SDSS spectrum, and an upper limit for the whole galaxy
from the H$\alpha$ equivalent width and the $r$-band flux. The photometric
rate is typically between the other two, and all of them increase
approximately linearly with galaxy luminosity. In principle, the past-average
SFR inferred from photometry and the current SFR inferred from H$\alpha$ may
be different, but in fact they are fairly similar.  The differences are
mostly the result of the different regions viewed.  The few galaxies with
H$\alpha$-fiber star formation rates that are less than their photometric
star formation rates have fiber positions that do not include the bright
heads. Generally, the stellar populations that dominate the heads were
produced at a rate that is comparable to the current H$\alpha$ rate.

\begin{figure}\epsscale{1}
%\plotone{f10newnew.jpg}
\plotone{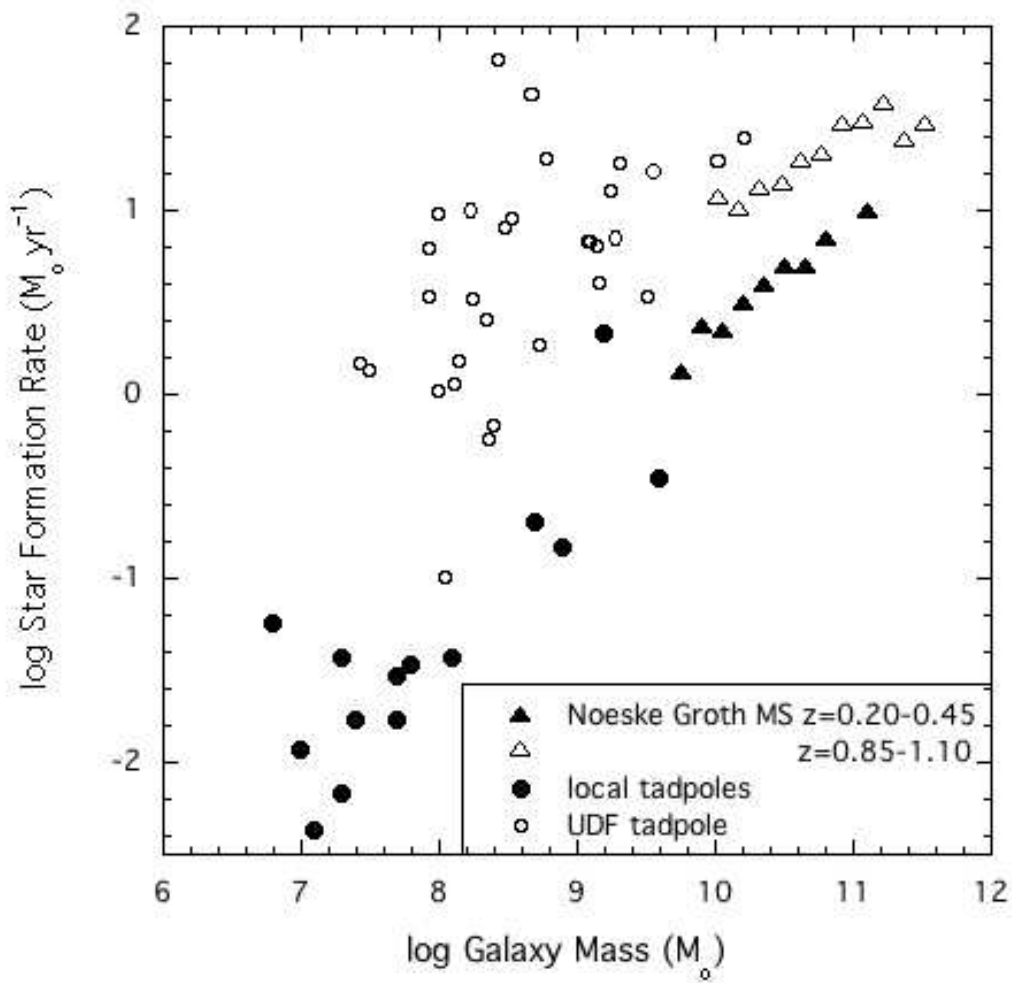}
\caption{The log of the star formation rate versus total stellar galaxy mass is shown
for the local tadpole galaxies (filled circles) compared with the ``main sequence''
Groth Strip galaxies in the redshift bin z=0.2 to 0.4 (filled triangles) from \cite{noeske07}.
The local tadpoles continue the relation of decreasing star
formation rate with decreasing galaxy mass. UDF tadpoles (open circles) from \cite{elm10} fit on the main sequence for the z=0.85 to 1.10 galaxies \citep{noeske07}. Higher redshift galaxies have higher star formation rates for a given mass.
\label{sfrm}}
\end{figure}

Many recent studies have examined the star formation rate in a large sample
of galaxies as a function of redshift. The star formation rate depends on the
stellar mass of the galaxy and its redshift. For example, \cite{noeske07}
compiled star formation rates for 2905 field galaxies in the Groth Strip from
GALEX and MIPS data and binned them into 4 redshift bins, from $z=0.2-0.45$
to $z=0.85-1.10$. They found increasing average rates for a given galaxy mass
with increasing redshift. \cite{bauer} examined star formation rates in 1300
GOODS galaxies binned from $z$=1.50 to 3.00 using HST UV and Spitzer IR data,
and found approximately constant rates over this redshift range that are
about 5 times higher than the star formation rates in the $z$=0.3 bin of
\cite{noeske07}. In Figure \ref{sfrm} we plot as filled circles the log of the star formation
rate versus total stellar galaxy mass for our local tadpole sample. The star
formation rates shown are those determined for the tadpole heads as listed in
Table \ref{tab4}. The galaxy masses are assumed to be the sum of the head
masses, the underlying disk masses that were subtracted from the head, and
the tail masses, also in Table \ref{tab4}. On the same figure we plot as filled triangles the
\cite{noeske07} $z=0.2-0.4$ average values in each mass bin. All but two of
the local galaxies fit well on this star formation main sequence (Kiso 3867
and 8466 have higher SFRs for their masses), indicating that the local
tadpoles have normal star formation rates for their masses. 

In Figure \ref{sfrm} we also show as open circles the UDF tadpoles from \cite{elm10}. The masses are taken to be the sum of the head and tail masses, and the star formation rates are assumed to be the sum of the head and tail rates. Plotted as open triangles are the $z=0.85-1.10$ average values from \cite{noeske07}. For both tadpoles and other galaxies, the star formation rates for a given mass are higher at higher redshift. The scatter in the \cite{noeske07} sample, as plotted in that paper, corresponds to about a factor of 10 in the star formation rate, the same as the scatter in our tadpole sample. The UDF tadpoles fit on this higher redshift main sequence, as the local tadpoles do for the nearby main sequence.

\begin{figure}\epsscale{1}
%\plotone{f11new.jpg}
\plotone{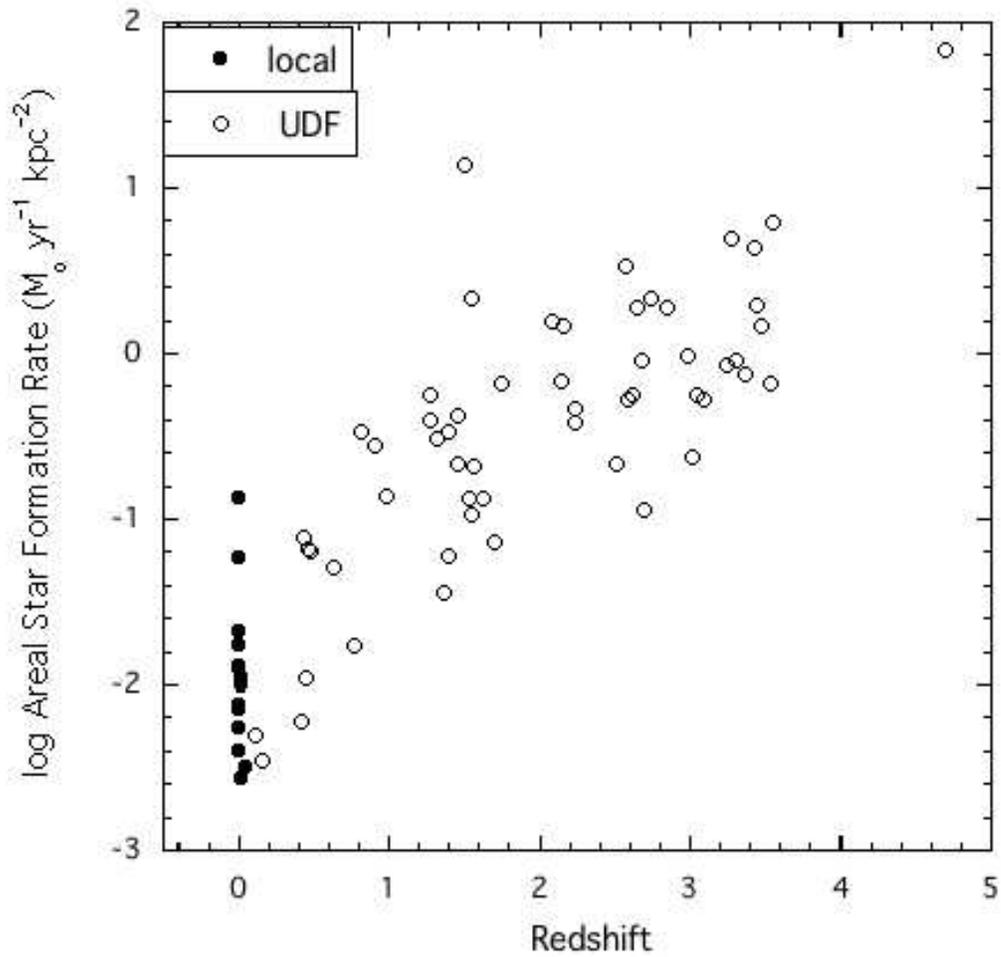}
 \caption{Areal star formation rate is shown
as a function of redshift for the heads of the local (black dots) and
high redshift UDF (open circles) tadpole galaxies. \cite{daddi10} show that starbursting galaxies have log areal rates greater than -1, so with this definition most of the UDF tadpoles are starbursting, whereas most of the local tadpoles are not. This dichotomy probably reflects more gas availability in the younger universe.}\label{ssfr}\end{figure}

Figure \ref{ssfr} shows the star formation rate per unit area, $\Sigma_{SFR}$
in $M_{\odot}$ yr$^{-1}$ kpc$^{-2}$, in the head region from photometry as a
function of redshift (also listed in Table \ref{tab4}). The local tadpoles,
shown as black dots, have a wide range of $\log\Sigma_{SFR}$ with a mean
value of $-2.0\pm0.47$. This is comparable to the areal star formation
rate of local spiral galaxies \citep{bigiel}.  The UDF tadpoles have higher
areal rates, increasing by about 2 orders of magnitude out to $z\sim2$. This
rise is faster than the universal star formation rate per unit volume as a
function of redshift shown by \cite{bouwens11}, which is $\sim$1.5 orders of
magnitude over the same redshift range. For comparison, the SINS sample of
galaxies \citep{genzel11} has seven clumps in five $z\sim2$ clumpy galaxies
with $\log\Sigma_{SFR}$ based on H$\alpha$ spectra ranging from $-0.15$ to
$1.1$. These values are slightly higher than the $\log\Sigma_{SFR}$ of the
UDF tadpole heads of similar size at the same redshift. \cite{daddi10} and
\cite{genzel10} study the areal star formation rate for different samples of
galaxies. \cite{daddi10} shows that the starburst galaxies have values
greater than about $0.1\;M_{\odot}$ yr$^{-1}$ kpc$^{-2}$, whereas normal
spirals lie below this value. In our Figure \ref{ssfr}, the local tadpoles
fall in the normal range while most of the high redshift tadpoles are in the
starburst range.

The specific star formation time, $\tau_{spec}$, in Gyr, is also given
in Table \ref{tab4}. This time equals the ratio of the total stellar
mass in the galaxy to the photometric star formation rate in the head.
%The total stellar mass was determined from the sum of the young-star head mass plus the underlying red-star %mass in the head, plus the tail mass.
The specific star formation times are a Gyr or more, which
suggests that these galaxies are not bursting at an unusually high
rate.

\section{Conclusions}
Tadpole galaxies in the Kiso and Michigan Surveys were studied as local
analogs of high redshift tadpoles. Characteristic of our local sample are
massive off-center star-forming clumps and diffuse tails, which could be
faint galaxy disks. The local tadpoles are extremely rare, amounting to only
 $\sim0.2$\% of UV-bright galaxies in the Kiso survey, while in the ACS
image of the UDF, tadpoles represent 10\% of all large galaxies
\citep{elm05}. Because of selection effects, the local tadpoles are less
massive than high redshift tadpoles, but their young-star head masses scale
with luminosity in the same way, ranging from $\sim10^5\;M_\odot$ to
$10^9\;M_{\odot}$ for restframe $U$-band absolute magnitudes of -13 mag to
-20 mag. The head masses increase to $10^{10}\; M_{\odot}$ in the most
luminous high redshift tadpoles. The star formation rate per unit area
increases with increasing redshift by 2 orders of magnitude from $z$=0 to 3.
The star formation rate per unit galaxy mass also increases with increasing redshift for the tadpoles, in agreement with the increase found previously for the Groth strip main sequence galaxies.

There are four indications from the present study that many tadpole galaxies
are normal faint disks with most of their star formation on one side.  One
indication is the similarity of the exponential scale length on each side of
the major axis in the far outer part, which suggests the presence of a faint
symmetric component. Another is the decrease in relative head mass over time,
which is an expected trend for normal galaxies with decreasing relative gas
abundance and increasing underlying star mass. A third indication is the
large value of the specific star formation time, i.e., a Gyr or more, which
implies a steady rate over the life of the galaxy.  A fourth indication is
the increasing age of the underlying disk relative to the star-forming head
with decreasing redshift. All of the tadpoles in this survey are of the type
having a blue head and a relatively red tail. The only other Kiso tadpole in
SDSS, KUG0010+371 (which has no spectrum or redshift information), may be a
counterexample, with a red head and a blue tail.

The origin of the lop-sided star formation is not understood. It could be
ram-pressure triggered as a result of galaxy motion though an intergalactic
medium, it could be the starburst reaction to an impact or galaxy
interaction, or it could be a random collapse of local disk gas with an
unstable Jeans length comparable to the galaxy radius. All of these origins are consistent with a lower fraction of tadpoles in the local universe than at high z, because there is less intergalactic gas, fewer collisions, and less galactic gas in today's universe.

\label{sect:conc} {\it Acknowledgments:} We thank the Vassar
Undergraduate Research Summer Institute (URSI) for support for JD and
the National Science Foundation for support for JP through the Keck
Northeast Astronomy Consortium from NSF REU grant AST-1005024. DME thanks Vassar College for page charge support. JSA and
CMT has been partly funded by the Spanish MICINN, project 'estallidos'
AYA~2007-67965-C03-01 and AYA~2010-21887-C04-04, and they are members
of the Consolider-Ingenio program 'first science with GTC', grant
MICINN CSD2006-00070. Funding for the Sloan Digital Sky Survey (SDSS)
has been provided by the Alfred P. Sloan Foundation, the Participating
Institutions, the National Aeronautics and Space Administration, the
National Science Foundation, the U.S. Department of Energy, the
Japanese Monbukagakusho, and the Max Planck Society. We also thank the
referee for helpful suggestions and references. The SDSS Web site is
http://www.sdss.org/. The SDSS is managed by the Astrophysical Research
Consortium (ARC) for the Participating Institutions. The Participating
Institutions are The University of Chicago, Fermilab, the Institute for
Advanced Study, the Japan Participation Group, The Johns Hopkins
University, Los Alamos National Laboratory, the Max-Planck-Institute
for Astronomy (MPIA), the Max-Planck-Institute for Astrophysics (MPA),
New Mexico State University, University of Pittsburgh, Princeton
University, the United States Naval Observatory, and the University of
Washington. This research has made use of the NASA/IPAC Extragalactic
Database (NED) which is operated by the Jet Propulsion Laboratory,
California Institute of Technology, under contract with the National
Aeronautics and Space Administration.

\clearpage

\begin{deluxetable}{llccc}
%table 1
\tabletypesize{\scriptsize}\tablecolumns{8} \tablewidth{0pt} \tablecaption{Galaxy properties}

\tablehead{\colhead{Galaxy\tablenotemark{a}}&\colhead{KUG no.,}& \colhead{Distance\tablenotemark{b}}&\colhead {R$_{25}$\tablenotemark{c}}&\colhead{M$_g$,tot\tablenotemark{c}}   \\
\colhead{}&\colhead{other}&\colhead{(Mpc)}&\colhead{(kpc)}&\colhead{(mag)}}

\startdata
3193    &   0852+400A   &   25.5    &   1.76    &   -15.32 \\
3473    &   0917+527,Mrk1416    &   32.9    &   3.93    &   -16.2   \\
3867    &   0937+298    &   7.4 &   1.38    &   -13.45   \\
3975    &   0940+544    &   22.8    &   2.81    &   -14.6      \\
5149    &   1113+237    &   172 &   22.9    &   -19.78    \\
5639    &   1138+327    &   24.5 &  3.35    &   -16.4    \\
5870    &   1149+224    &   46.9    &   4.34    &   -16.57   \\
6511    &   1220+124,IC3224 &   16.8    &   1.78    &   -14.85   \\
6610    &   1225+253,IC3384 &   18.1    &   2.46    &   -14.59   \\
6664    &   1229+151,IC3453,Mrk1328 &   34.9    &   12.2    &   -18.02   \\
6669    &   1229+276,IC3460 &   61.6    &   18.0  &   -17.97  \\
6877    &   1243+265    &   26.3    &   2.51    &   -15.65   \\
8466    &   1601+192,Mrk296 &   64.5    &   13.5    &   -18.78   \\
UM 417  &    -  &   37.3    &   2.97    &   -14.85   \\
\label{tab1}
\enddata
\tablenotetext{a}{Kiso catalog number except last entry, which is a University of Michigan number}
\tablenotetext{b}{From SDSS-DR7 redshift using H$_o$=73 except for Kiso 3975, 6511, and UM 417, whose distances are from  NASA/IPAC Extragalactic Database, http://ned.ipac.caltech.edu}
\tablenotetext{c}{From photometric measurements of SDSS images and radial profiles along the major axis}

\end{deluxetable}
\clearpage
%table 2
\begin{deluxetable}{lccccccc}
\tabletypesize{\scriptsize}\tablecolumns{6} \tablewidth{0pt}
\tablecaption{Surface Brightness Profiles}

\tablehead{
\colhead{Galaxy}&
\colhead{$r_s$\tablenotemark{a}}&
\colhead{$r_s/$$R_{25}$}&
\colhead{No.\tablenotemark{b}} &
\colhead{left $r_s$\tablenotemark{c}}  &
\colhead{No.\tablenotemark{b} }  &
\colhead{right $r_s$\tablenotemark{d}}  &
\colhead{No.\tablenotemark{b}}\\
\colhead{}&
\colhead{(kpc)}&
\colhead{}&
\colhead{}    &
\colhead{(kpc)}  &
\colhead{} &
\colhead{(kpc)}  &
\colhead{}}
\startdata

3193    &   0.89    $\pm$   0.036   &   0.51    &   3.4 &   0.88    $\pm$   0.05    &   2.5 &   1.04    $\pm$   0.21    &   1.4 \\
3473    &   2.14    $\pm$   0.011   &   0.55    &   2.1 &   0.80    $\pm$   0.02    &   4.9 &   0.7 $\pm$   0.29    &   1   \\
3867    &   0.98    $\pm$   0.018   &   0.71    &   2.8 &   1.06    $\pm$   0.08    &   1.9 &   0.95    $\pm$   0.02    &   3.3 \\
3975    &   3.18    $\pm$   0.076   &   0.81    &   1.3 &   0.64    $\pm$   0.06    &   2.2 &   0.28    $\pm$   1.71    &   5.4 \\
5149    &   -           &   -   &    -  &   5.67    $\pm$   0.26    &   2.5 &   4.05    $\pm$   0.11    &   3.6 \\
5639    &   5.71    $\pm$   21      &   1.70    &   0.9 &   0.44    $\pm$   0.52    &   2.1 &   0.88    $\pm$   0.84    &   2.5 \\
5870    &   4.94    $\pm$   0.22    &   1.14    &   2.6 &   3.1 $\pm$   0.61    &   1.4 &   1.6 $\pm$   0.14    &   2.4 \\
6511    &   3.00    $\pm$   0.88    &   1.68    &   0.5 &   0.82    $\pm$  0.07  &   2.9 &   0.45    $\pm$   0.03    &   2.9 \\
6610    &   11.8    $\pm$   3.43    &   4.81    &   0.4 &   1.00    $\pm$   0.09    &   1.3 &   0.63    $\pm$   0.03    &   1.6 \\
6664    &   22.4    $\pm$   2.62    &   1.84    &   0.7 &   4.74    $\pm$   0.19    &   1.5 &   3.87    $\pm$   0.14    &   1.5 \\
6669    &   18.3    $\pm$   0.88    &   1.02    &   0.8 &   6.09    $\pm$   1.2 &   1.9 &   2.31    $\pm$   1.8 &   2.8 \\
6877    &   2.16    $\pm$   0.15    &   0.86    &   1.2 &   0.44    $\pm$   0.04    &   3.6 &   0.71    $\pm$   0.06    &   2.7 \\
8466    &   264.8   $\pm$   62  &   89.2    &   0.03    &   2.87    $\pm$   0.12    &   2.6 &   4.1 $\pm$   0.18    &   2.5 \\
UM 417   &   101 $\pm$   1899    &   34  &   0.04    &   0.45    $\pm$   0.04    &   5.1 &   0.77    $\pm$   0.18    &   3.2 \\
    \label{tab2}
\enddata
\tablenotetext{a}{Scale length measured from the relatively flat part of the
surface brightness profile} \tablenotetext{b}{Number of scale lengths that
fit in linear part of profile where measured} \tablenotetext{c}{Scale length
measured from outer eastern surface brightness profile}
\tablenotetext{d}{Scale length measured from outer western surface brightness
profile}
\end{deluxetable}

\clearpage

\begin{deluxetable}{lccccccc}
%new table 3
\tabletypesize{\scriptsize}\tablecolumns{8} \tablewidth{0pt}
\tablecaption{Masses, Ages, and Surface Densities}

\tablehead{
\colhead{Galaxy\tablenotemark{a}}&
\colhead{log $M_{gal}$\tablenotemark{b}}&
\colhead{log M$_{Head}$\tablenotemark{b} }&
\colhead{log M$_{Tail}$\tablenotemark{b}}&
\colhead{log Age$_{Head}$\tablenotemark{b}}&
\colhead{log Age$_{Tail}$\tablenotemark{b}}&
\colhead{$\Sigma_{Head}$\tablenotemark{c}}&
\colhead{ $\Sigma_{Tail}$\tablenotemark{c}}  \\
\colhead{}&
\colhead{($M_{\odot}$)}&
\colhead{($M_{\odot}$)}&
\colhead{($M_{\odot}$)}&
\colhead{(yr)}&
\colhead{(yr)}&
\colhead{($M_{\odot}$ pc$^{-2}$)}&
\colhead{($M_{\odot}$ pc$^{-2}$)}}
\startdata
3193  & 7.3 $\pm$ 0.4  &   6.4 $\pm$   0.1 &   7.1 $\pm$   0.1 & 8.5 $\pm$ 0.2 & 9.4 $\pm$ 0.4 &2.4  &   5.1 \\
3473  & 7.8 $\pm$ 0.4  &   6.9 $\pm$   0.1 &   7.5 $\pm$   0.1 & 8.3 $\pm$ 0.3 & 9.4 $\pm$ 0.3 &1.2  &   3.9 \\
3867  & 6.8 $\pm$ 0.4  &   5.2 $\pm$   0.3 &   6.7 $\pm$   0.1 & 6.4 $\pm$ 0.7 & 9.4 $\pm$ 0.3 &0.35 &   2.8 \\
3975  & 7.0 $\pm$ 0.4  &   5.5 $\pm$   0.3 &   7.0 $\pm$   0.1 & 7.4 $\pm$ 0.6 & 9.2 $\pm$ 0.3 &0.45 &   0.85    \\
5149  & 9.6 $\pm$ 0.4  &   8.9 $\pm$   0.1 &   9.5 $\pm$   0.1 & 9.3 $\pm$ 0.4 & 9.4 $\pm$ 0.3 &6.6  &   4.5 \\
5639  & 7.7 $\pm$ 0.4  &   6.7 $\pm$   0.2 &   7.6 $\pm$   0.1 & 8.5 $\pm$ 0.4 & 9.1 $\pm$ 0.2 &2.5  &   3.4 \\
5870  & 8.1 $\pm$ 0.4  &   6.1 $\pm$   0.2 &   8.0 $\pm$   0.2 & 7.5 $\pm$ 0.4 & 9.4 $\pm$ 0.4 &0.09 &   1.9 \\
6511  & 7.7 $\pm$ 0.4  &   5.8 $\pm$   0.1 &   7.6 $\pm$   0.1 & 7.3 $\pm$ 0.3 & 9.5 $\pm$ 0.3 &0.28 &   4.6 \\
6610  & 7.4 $\pm$ 0.7  &   5.1 $\pm$   0.7 &   7.4 $\pm$   0.1 & 6.8 $\pm$ 1.3 & 9.5 $\pm$ 0.3 &0.40 &   2.6 \\
6664  & 8.7 $\pm$ 0.4  &   6.7 $\pm$   0.1 &   8.7 $\pm$   0.1 & 7.4 $\pm$ 0.4 & 9.3 $\pm$ 0.4 &0.57 &   4.1 \\
6669  & 8.9 $\pm$ 0.4  &   7.0 $\pm$   0.1 &   8.8 $\pm$   0.1 & 7.8 $\pm$ 0.5 & 9.3 $\pm$ 0.4 &0.77 &   2.1 \\
6877  & 7.3 $\pm$ 0.4  &   6.3 $\pm$   0.1 &   6.7 $\pm$   0.1 & 7.7 $\pm$ 0.2 & 9.4 $\pm$ 0.3 &0.65 &   1.9 \\
8466  & 9.2 $\pm$ 0.3  &   8.1 $\pm$   0.1 &   8.9 $\pm$   0.1 & 7.8 $\pm$ 0.6 & 9.4 $\pm$ 0.3 &0.62 &   3.8 \\
UM 417 & 7.1 $\pm$ 0.4 &   6.4 $\pm$   0.1 &   7.0 $\pm$   0.2 & 8.7 $\pm$ 0.2 & 7.1 $\pm$ 1.5\tablenotemark{d}  &2.1 &   0.5 \\
\label{tab3}
\enddata
\tablenotetext{a}{Kiso catalog number except last entry, which is a University of Michigan number}
\tablenotetext{b}{From fits of photometric measurements to evolutionary models, as described in text. An underlying disk component was subtracted from the heads, and a metallicity of 0.4 solar was assumed.}
\tablenotetext{c}{Surface density, from mass and area, as described in text}
\tablenotetext{d}{For 0.02 times solar metallicity, the age of the tail is more like the other ages, 9.3$\pm$0.4 in the log}
\end{deluxetable}
\clearpage
\begin{deluxetable}{lcccccc}
\tabletypesize{\scriptsize}\tablecolumns{6} \tablewidth{0pt} \tablecaption{Star Formation Rates}

\tablehead{\colhead{Galaxy}& \colhead{log SFR$_{Head}$\tablenotemark{a}}&\colhead{log SFR$_{Tail}$\tablenotemark{b}}&
\colhead{log area SFR$_{Head}$\tablenotemark{c}} & \colhead{log SFR spec\tablenotemark{d}} &
\colhead{log SFR mag\tablenotemark{e}}&\colhead{$\tau_{spec}$\tablenotemark{f}}  \\
\colhead{}&\colhead{($M_{\odot}$ yr$^{-1}$)}&\colhead{($M_{\odot}$
yr$^{-1}$)}&\colhead{($M_{\odot}$ yr$^{-1}$
kpc$^{-2}$)}&\colhead{($M_{\odot}$ yr$^{-1}$)}&\colhead{($M_{\odot}$
yr$^{-1}$)}&\colhead{Gyr}} \startdata

3193    &   -2.17                &   -2.25   &   -2.16   &   -1.71   &   -0.91 & 3.07 \\
3473    &   -1.47                &   -1.95   &   -2.26   &   -1.07   &   -0.15 & 1.82 \\
3867    &   -1.25                &   -2.70   &   -0.87   &   -3.01   &   -2.05 & 0.11 \\
3975    &   -1.93                &   -2.24   &   -1.76   &   -2.94   &   -1.94 & 0.89 \\
5149    &   -0.46                &    0.02   &   -2.50   &   -0.26   &    0.28 & 12.2 \\
5639    &   -1.77                &   -1.50   &   -2.12   &   -1.15   &   -0.19 & 2.99 \\
5870    &   -1.43                &   -1.44   &   -2.57   &   -1.83   &   -0.65 & 3.16 \\
6511    &   -1.54                &   -1.88   &   -1.89   &   -2.12   &   -0.80 & 1.88 \\
6610\tablenotemark{g}  &   -1.77 &   -2.13   &   -1.23   &   -3.54   &   -2.24 & 1.53 \\
6664    &   -0.70                &   -0.63   &   -1.68   &   -2.70   &   -1.02 & 2.77 \\
6669    &   -0.84                &   -0.60   &   -1.95   &   -1.64   &   -0.32 & 4.91 \\
6877    &   -1.44                &   -2.71   &   -1.90   &   -1.93   &   -0.96 & 0.51 \\
8466    &   0.33                 &   -0.52   &   -2.00   &   -0.83   &    0.23 & 0.83 \\
UM 417\tablenotemark{h}&   -2.37 &   -0.16   &   -2.41   &       -   &    -    & 2.82 \\
    \label{tab4}
\enddata
\tablenotetext{a}{Mean star formation rate for head based on mass and age
from model fits to photometry} \tablenotetext{b}{Mean star formation rate for
tail based on mass and age from model fits to photometry}
\tablenotetext{c}{Mean star formation rate per unit area for head}
\tablenotetext{d} {Current star formation rate for whole galaxy based on
central 3'' SDSS H$\alpha$ spectrum} \tablenotetext{e} {Current star
formation rate for whole galaxy based H$\alpha$ flux estimated from galaxy
$r$-band magnitude} \tablenotetext{f} {Specific star formation time,
determined from galaxy mass divided by head star formation rate, where galaxy
mass is the sum of the head and tail masses plus the tail surface density
times the head area (that is, the mass presumed to be underlying the head).}
\tablenotetext{g}{Brightest two sub-clumps used for head mass}
\tablenotetext{h}{No SDSS spectrum}
\end{deluxetable}

\clearpage
%fig1
 \clearpage

%fig2
\clearpage

%fig3
 \clearpage

%fig4
\clearpage
 \clearpage

%fig5
\clearpage
 \clearpage

%fig6
 \clearpage

%fig7
\clearpage
 \clearpage

%fig8
 \clearpage

%fig9
 \clearpage

%fig10

%fig11
\clearpage
\clearpage \clearpage

\end{document}